\documentclass[11pt,a4paper]{article}
\usepackage{amsfonts,amsmath,amssymb,bm,cite}

\begin{document}

\title{\bf Neutrino spin-flavor oscillations \\ in electromagnetic fields of 
various configurations}
\author{Maxim Dvornikov$^{a,b}$\footnote{{\bf e-mail}: dvmaxim@cc.jyu.fi}
\\
$^a$ \small{\em University of Jyv\"{a}skyl\"{a}} \\
\small{\em Department of Physics, P.O. Box 35, FIN-40014, Finland;}
\\
$^b$ \small{\em Institute of Terrestrial Magnetism, 
Ionosphere and Radiowave Propagation} \\
\small{\em 142190, Troitsk, Moscow region, Russia}
}
\date{}
\maketitle

\begin{abstract}
We study spin-flavor oscillations of Dirac neutrinos with mixing and having 
non-zero matrix of magnetic moments in magnetic fields of various 
configurations. We discuss constant transversal and twisting magnetic fields. 
To describe the dynamics of Dirac neutrinos we use relativistic quantum 
mechanics approach based on the exact solutions to the Dirac-Pauli equation in 
an external electromagnetic field. We derive transition probabilities for 
different neutrino magnetic moments matrices.  
\end{abstract}

\section{Introduction}

Neutrino conversions from one flavor to another combined with the change of the 
particle helicity, e.g. $\nu_e^\mathrm{L} \leftrightarrow \nu_\mu^\mathrm{R}$, 
are usually called neutrino spin-flavor oscillations (see Refs.~\cite{qmSFO}). 
This neutrino oscillations type is important since it could be one of the 
possible explanations of the time variability of the solar neutrino flux (see, 
e.g., Refs.~\cite{solarnu}). However it was suggested in 
Refs.~\cite{smallcontr} that neutrino spin-flavor oscillations in solar 
magnetic fields give a sub-dominant contribution in to the total conversion of 
solar neutrinos.

In this paper we summarize the results of our recent studies (see 
Refs.~\cite{DvoMaa07,Dvo07}) of neutrino spin-flavor oscillations in external 
electromagnetic fields of various configurations. We suppose that neutrinos are 
Dirac particles. To describe the evolution of the neutrino system we use the 
approach based on the relativistic quantum mechanics. We start from exact 
solutions to the Dirac-Pauli equation in an external magnetic field and then 
derive the neutrino wave functions satisfying the given initial condition. We 
used this method to describe neutrino flavor and spin-flavor oscillations in 
vacuum and in various external fields (see 
Refs.~\cite{FOvac,Dvo06EPJC,DvoMaa07,Dvo07}).
Note that neutrino neutrino spin-flavor oscillations in electromagnetic fields 
of various configurations were examined in Refs.~\cite{emfields} using the 
standard quantum mechanical approach. The propagation and oscillations of 
neutrinos in strong magnetic fields was also studied in Refs.~\cite{strongB}. 

In Sec.~\ref{ELECTRODYN} we formulate the initial condition problem for the 
system of two Dirac neutrinos which mix and have non-vanishing matrix of 
magnetic moments. Moreover the mass and the magnetic moments matrices are 
assumed to independent since we study this system on the phenomenological 
level. It means that the diagonalization of the mass matrix does not involve 
the diagonalization of the magnetic moments matrix. In Sec.~\ref{CONSTB} on the 
basis of the known solution to the Dirac-Pauli equation in the constant 
transversal magnetic field we describe the time evolution of the system in 
question. We obtain the most general final neutrino wave function which exactly 
takes into account all neutrino magnetic moments and valid for arbitrary 
strength of the external magnetic field. Then we discuss several applications 
of the yielded results and calculate the transition probability for various 
magnetic moments matrices. In Sec.~\ref{TWISTB} we apply the same technique for 
the analysis of spin-flavor oscillations of Dirac neutrinos in the twisting 
magnetic field. We start from the recently obtained solution to the Dirac-Pauli 
equation for this magnetic field configuration. We derive transition 
probabilities for various types of neutrino magnetic moments matrices. Then in 
Sec.~\ref{CONCL} we summarize our results.

\section{Electrodynamics of mixed neutrinos with magnetic 
moments}\label{ELECTRODYN}

Let us study the evolution of two Dirac neutrinos $(\nu_\alpha,\nu_\beta)$ that 
mix and interact with the external electromagnetic field 
$F_{\mu\nu}=(\mathbf{E},\mathbf{B})$. The Lagrangian for this system has the 
form
\begin{equation}\label{Lagrnu}
  \mathcal{L}(\nu_{\alpha},\nu_{\beta})=
  \sum_{\lambda=\alpha,\beta}\bar{\nu}_\lambda 
  \mathrm{i}\gamma^\mu\partial_\mu \nu_\lambda-  
  \sum_{\lambda\lambda'=\alpha,\beta}
  \left[
    m_{\lambda\lambda'} \bar{\nu}_\lambda \nu_{\lambda'}+
	\frac{1}{2}	 
	M_{\lambda\lambda'}
	\bar{\nu}_{\lambda}\sigma_{\mu\nu}\nu_{\lambda'} F^{\mu\nu}
  \right],
\end{equation}
where 
$\sigma_{\mu\nu}=(\mathrm{i}/2)(\gamma_\mu\gamma_\nu-\gamma_\nu\gamma_\mu)$. 
The neutrino mass matrix $(m_{\lambda\lambda'})$ and the neutrino magnetic 
moments matrix $(M_{\lambda\lambda'})$ are generally independent.

To describe the dynamics of the system we set the initial condition by 
specifying the initial wave functions of flavor neutrinos $\nu_{\lambda}$ and 
then analytically determine the wave functions at subsequent moments of time. 
We assume that the initial condition is
\begin{equation}\label{inicondnu}
  \nu_{\alpha}(\mathbf{r},0)=0,
  \quad
  \nu_{\beta}(\mathbf{r},0)=\xi(\mathbf{r}),
\end{equation}
where $\xi(\mathbf{r})$ is a given function. Let us choose it in the following 
form: $\xi(\mathbf{r})=e^{\mathrm{i}\mathbf{k}\mathbf{r}}\xi_0$, where 
$\mathbf{k}=(k,0,0)$ and $\xi_0^\mathrm{T}=(1/2)(1, -1, -1, 1)$. One can check 
that $(1/2)(1-\Sigma_1)\xi_0=\xi_0$, i.e. the wave function $\xi(\mathbf{r})$ 
corresponds to a left-handed relativistic neutrino propagating along the 
$x$-axis. The similar choice of the initial condition was adopted in our 
works~\cite{FOvac,Dvo06EPJC,DvoMaa07,Dvo07}.

In order to diagonalize the mass matrix in Eq.~\eqref{Lagrnu} we introduce  the 
mass eigenstates wave functions, $\psi_a$, $a=1,2$, obtained from the original 
flavor wave functions $\nu_{\lambda}$ through the unitary transformation
\begin{equation}\label{matrtrans}
  \nu_{\lambda}=\sum_{a=1,2}U_{\lambda a}\psi_a,
\end{equation}
where the matrix $({U}_{\lambda a})$ is parametrized with help of the mixing 
angle $\theta$,
\begin{equation}\label{matrU}
  ({U}_{\lambda a})=
  \begin{pmatrix}
    \cos \theta & -\sin \theta \\
    \sin \theta & \cos \theta \
  \end{pmatrix}.
\end{equation}
The Lagrangian~\eqref{Lagrnu} rewritten in terms of the fields $\psi_a$ takes 
the form
\begin{equation}\label{Lagrpsi}
  \mathcal{L}(\psi_1,\psi_2)=\sum_{a=1,2}
  \bar{\psi}_a(\mathrm{i}\gamma^\mu \partial_\mu-m_a)\psi_a-
  \frac{1}{2}
  \sum_{ab=1,2}\mu_{ab}\bar{\psi}_a\sigma_{\mu\nu}\psi_b F^{\mu\nu},
\end{equation}
where $m_a$ is the mass of the fermion $\psi_a$ and
\begin{equation}\label{magmomme}
  \mu_{ab}=\sum_{\lambda\lambda'=\alpha,\beta}
  U^{-1}_{a\lambda}{M}_{\lambda\lambda'}U_{\lambda' b},
\end{equation}
is the magnetic moment matrix presented in the mass eigenstates basis. Note 
that the matrix $(\mu_{ab})$ in Eq.~\eqref{magmomme} can be non-diagonal, i.e. 
the transition magnetic moment can have non-zero value, $\mu_{12}=\mu_{21}=\mu 
\neq 0$. 

Let us assume that the electric field vanishes, $\mathbf{E}=0$. In this case we 
write down the Dirac-Pauli equation for $\psi_a$, resulting from 
Eq.~\eqref{Lagrpsi}, as follows:
\begin{equation}\label{Direqpsi}
  \mathrm{i}\dot{\psi}_a=\mathcal{H}_a\psi_a+V\psi_b,
  \quad
  a,b=1,2,
  \quad
  a \neq b,
\end{equation}
where $\mathcal{H}_a=(\bm{\alpha}\mathbf{p})+\beta m_a-\mu_a \beta 
(\bm{\Sigma}\mathbf{B})$ is the Hamiltonian for the particle $\psi_a$ 
accounting for the magnetic field, $V=-\mu \beta (\bm{\Sigma}\mathbf{B})$ 
describes the interaction of the transition magnetic moment with the external 
magnetic field and $\mu_a=\mu_{aa}$.  

The general solution to Eq.~\eqref{Direqpsi} can be presented as follows:
\begin{equation}\label{GsolDPeq}
  \psi_{a}(\mathbf{r},t)=
  \int \frac{\mathrm{d}^3\mathbf{p}}{(2\pi)^{3/2}}
  e^{\mathrm{i}\mathbf{p}\mathbf{r}}\sum_{\zeta=\pm 1}
  \left[
    a_a^{(\zeta)}(t)u_a^{(\zeta)}\exp{(-\mathrm{i}E_a^{(\zeta)} t)}+
    b_a^{(\zeta)}(t)v_a^{(\zeta)}\exp{(+\mathrm{i}E_a^{(\zeta)} t)}
  \right],
\end{equation}
The basis spinors $u_a^{(\zeta)}$ and $v_a^{(\zeta)}$, as well as the energy 
$E_a^{(\zeta)}$, are the exact solutions to the Dirac equations,
\begin{equation}
  \mathcal{H}_a u_a^{(\zeta)} = E_a^{(\zeta)} u_a^{(\zeta)},
  \quad
  \mathcal{H}_a v_a^{(\zeta)} = - E_a^{(\zeta)} v_a^{(\zeta)},
\end{equation}
accounting for the external magnetic field. The discrete quantum number 
$\zeta=\pm 1$ describes different polarization states of the fermion $\psi_a$. 
The coefficients $a_a^{(\zeta)}$ and $b_a^{(\zeta)}$ in Eq.~\eqref{GsolDPeq} 
are in general functions of time.

\section{Neutrino spin-flavor oscillations in the transversal magnetic 
field}\label{CONSTB}

In this section we study the evolution of the system in question under the 
influence of the constant magnetic field directed along the $z$-axis, 
$\mathbf{B}=(0,0,B)$. 

The basis spinors and energy levels in Eq.~\eqref{GsolDPeq} can be found in 
Refs.~\cite{TerBagKha65,DvoMaa07}. The energy as a function of the particle 
mass and momentum $\mathbf{p}=(p_1,p_2,p_3)$ is
\begin{equation}\label{energyconstB}
  E_a^{(\zeta)} = \sqrt{p_3^2+\mathcal{E}_a^{(\zeta)2}},
  \quad
  \mathcal{E}_a^{(\zeta)}=\mathcal{E}_a-\zeta \mu_a B,
  \quad
  \mathcal{E}_a = \sqrt{m_a^2+p_1^2+p_2^2}.
\end{equation}
The basis spinors are expressed in the following form:
\begin{equation}\label{spinorsconstB}
  u_a^{(\zeta)}=
  \frac{1}{2\sqrt{E_a^{(\zeta)}}}
  \begin{pmatrix}
     \phi^{+{}}_a \alpha^{+{}}_a \\
     -\zeta\phi^{-{}}_a \alpha^{-{}}_a e^{\mathrm{i}\varphi} \\
     \phi^{+{}}_a \alpha^{-{}}_a \\
     \zeta\phi^{-{}}_a \alpha^{+{}}_a e^{\mathrm{i}\varphi} \
  \end{pmatrix},
  \quad
  v_a^{(\zeta)}=
  \frac{1}{2\sqrt{E_a^{(\zeta)}}}
  \begin{pmatrix}
     \phi^{+{}}_a \alpha^{-{}}_a \\
     \zeta\phi^{-{}}_a \alpha^{+{}}_a e^{\mathrm{i}\varphi} \\
     -\phi^{+{}}_a \alpha^{+{}}_a \\
     \zeta\phi^{-{}}_a \alpha^{-{}}_a e^{\mathrm{i}\varphi} \
  \end{pmatrix},
\end{equation}
where
\begin{equation*}
  \phi^{\pm{}}_a=\sqrt{1\pm\zeta m_a/\mathcal{E}_a},
  \quad
  \alpha^{\pm{}}_a=\sqrt{E_a^{(\zeta)}\pm\zeta\mathcal{E}_a^{(\zeta)}},
\end{equation*}
and $\tan \varphi = p_2/p_1$.

Using the results of our paper~\cite{DvoMaa07} we obtain the right-handed 
component of the wave function $\nu_\alpha$ accounting for the initial 
condition~\eqref{inicondnu},
\begin{align}\label{nualphaconstB}
  \nu_\alpha^{\mathrm{R}}(x,t)= &
  \bigg\{
    \sin\theta\cos\theta
    \frac{1}{2\mathrm{i}}
    \left[
      \frac{\omega_{+{}}}{\Omega_{+{}}}\sin(\Omega_{+{}}t)
      \exp{(\mathrm{i}\bar{\mu}Bt)}-
      \frac{\omega_{-{}}}{\Omega_{-{}}}\sin(\Omega_{-{}}t)
      \exp{(-\mathrm{i}\bar{\mu}Bt)}
    \right]
    \notag
    \\
    & +
    \mathrm{i}\mu B
    \left[
      \frac{\sin(\Omega_{+{}}t)}{\Omega_{+{}}}\cos^2\theta-
      \frac{\sin(\Omega_{-{}}t)}{\Omega_{-{}}}\sin^2\theta
    \right]
    \cos(\bar{\mu}Bt)
  \bigg\}
  \exp{(-\mathrm{i}\bar{\mathcal{E}}t+\mathrm{i}kx)}\kappa_0,
\end{align}
where $\Omega_{\pm{}}=\sqrt{(\mu B)^2+(\omega_{\pm{}}/2)^2}$, 
$\omega_{\pm{}}=E_1^{\pm{}}-E_2^{\pm{}}$, $\bar{\mu}=(\mu_1+\mu_2)/2$, 
$\bar{\mathcal{E}}=(\mathcal{E}_1+\mathcal{E}_2)/2$ and 
$\kappa_0^\mathrm{T}=(1/2)(1,1,1,1)$. Note that to obtain 
Eq.~\eqref{nualphaconstB} we approach to the relativistic limit $k \gg m_{a}$. 
Eq.~\eqref{nualphaconstB} is the most general one which accounts for all 
neutrino magnetic moments. The transition probability for the process 
$\nu_\beta^\mathrm{L} \to \nu_\alpha^\mathrm{R}$ can be calculated as 
$P_{\nu_\beta^\mathrm{L} \to 
\nu_\alpha^\mathrm{R}}(t)=|\nu_\alpha^{\mathrm{R}}(x,t)|^2$. It should be also 
mentioned that Eq.~\eqref{nualphaconstB} corresponds to neutrino spin-flavor 
oscillations in the transversal magnetic field since $\mathbf{k}=(k,0,0)$ and 
$\mathbf{B}=(0,0,B)$. 

Let us discuss two applications of Eq.~\eqref{nualphaconstB}. First we consider 
the case when $\mu_a \gg \mu$, i.e. the magnetic moments matrix in 
Eq.~\eqref{magmomme} is close to diagonal. In this situation the transition 
probability calculated from Eq.~\eqref{nualphaconstB} is
\begin{equation}\label{Ptr0constBD}
  P_{\nu_\beta^\mathrm{L}\to\nu_\alpha^\mathrm{R}}(t)= 
  \sin^2(2\theta)
  \left\{
    \sin^2(\delta\mu B t)\cos^2(\bar\mu B t)+
    \sin(\mu_1 B t)\sin(\mu_2 B t)
    \sin^2
    \left[
      \Phi(k)t
    \right]
  \right\},
\end{equation}    
where $\delta\mu = (\mu_1-\mu_2)/2$, $\Phi(k) = \delta m^2 / (4k)$ is the phase 
of vacuum oscillations and $\delta m^2 = m_1^2 - m_2^2$. In 
Eq.~\eqref{Ptr0constBD} we present the zero order (in $\mu$) contribution to 
the transition probability. The next order correction can be found in our 
work~\cite{DvoMaa07}. 

Now we consider the situation when $\mu \gg \mu_a$, i.e. magnetic moments 
matrix in Eq.~\eqref{magmomme} with great non-diagonal elements. In this case 
the transition probability based on Eq.~\eqref{nualphaconstB} is (see 
Ref.~\cite{DvoMaa07}),
\begin{equation}\label{PtrconstBM}
  P_{\nu_\beta^\mathrm{L}\to\nu_\alpha^\mathrm{R}}(t)=
  \cos^2(2\theta)
  \left(
    \frac{\mu B}{\Omega}
  \right)^2
  \sin^2(\Omega t),
\end{equation}
where $\Omega=\sqrt{(\mu B)^2+\Phi^2(k)}$.

Using Eqs.~\eqref{Ptr0constBD} and~\eqref{PtrconstBM} one can compute 
transition probabilities for spin-flavor oscillations of Dirac neutrinos when 
particles interact with the constant transversal magnetic field. However 
neutrino oscillations with the most general magnetic moments matrix should be 
studied on the basis of Eq.~\eqref{nualphaconstB}. 

\section{Neutrino spin-flavor oscillations in the twisting magnetic 
field}\label{TWISTB}

In this section we examine the evolution of two Dirac neutrinos under the 
influence of the twisting magnetic field, $\mathbf{B}=B(0,\sin \omega x,\cos 
\omega x)$, where $\omega$ is the frequency of the magnetic field rotation in 
space. Note that neutrino spin-flavor oscillations in the twisting magnetic 
field were studied in Refs.~\cite{twisting} in frames of the standard quantum 
mechanical approach. 

For this configuration of the magnetic field we should consider the modified 
wave functions in Eq.~\eqref{GsolDPeq}, $\psi_a \to \tilde{\psi}_a = 
\mathcal{U}^\dag \psi_a$, where 
$\mathcal{U}=\mathrm{diag}(\mathfrak{U},\mathfrak{U})$ and 
$\mathfrak{U}=\cos(\omega x/2)+\mathrm{i}\sigma_1\sin(\omega x/2)$. The 
Hamiltonian for the fermions $\tilde{\psi}_a$ appears to be $x$-coordinate 
independent (see Ref.~\cite{Dvo07}). The basis spinors and energy levels in the 
modified equation~\eqref{GsolDPeq} were found in Ref.~\cite{Dvo07}. The energy 
as a function of the particle mass and momentum, which is directed along the 
$x$-axis, $\mathbf{p}=(p,0,0)$, is
\begin{equation}\label{energytwistingB}
  E_a^{(\zeta)}=
  \sqrt{\mathcal{M}_a^2 + m_a^2 + p^2 - 2 \zeta R_a^2},
\end{equation}
where $R_a^2=\sqrt{p^2 \mathcal{M}_a^2 + (\mu_a B)^2 m_a^2}$ and 
$\mathcal{M}_a=\sqrt{(\mu_a B)^2 + \omega^2/4}$. The basis spinors take the 
following form in the relativistic limit:
\begin{align}\label{spinorstwistingB}
  u_a^{(\zeta)}= &
  \frac{1}{2\sqrt{2\mathcal{M}_a[\mathcal{M}_a+\zeta\omega/2]}}
  \begin{pmatrix}
     \mu_a B+\zeta\mathcal{M}_a+\omega/2 \\
     \mu_a B-\zeta\mathcal{M}_a-\omega/2 \\
     \mu_a B-\zeta\mathcal{M}_a-\omega/2 \\
     \mu_a B+\zeta\mathcal{M}_a+\omega/2 \
  \end{pmatrix},
  \notag
  \\
  v_a^{(\zeta)}= &
  \frac{1}{2\sqrt{2\mathcal{M}_a[\mathcal{M}_a-\zeta\omega/2]}}
  \begin{pmatrix}
     \mathcal{M}_a-\zeta \omega/2-\zeta \mu_a B \\
     \mathcal{M}_a-\zeta \omega/2+\zeta \mu_a B \\
     \zeta \omega/2-\mathcal{M}_a-\zeta \mu_a B \\
     \zeta \omega/2-\mathcal{M}_a+\zeta \mu_a B \
  \end{pmatrix}.
\end{align}
Note that the spinors in Eq.~\eqref{spinorstwistingB} satisfy the 
orthonormality conditions. 

First let us study the situation when the magnetic moments matrix in 
Eq.~\eqref{magmomme} is close to diagonal, i.e. $\mu_a \gg \mu$. Using the 
calculations analogous to those in Sec.~\ref{CONSTB} and the results of our 
work~\cite{Dvo07} we receive the zero order term (in $\mu$) in the expansion of 
the transition probability as
\begin{align}\label{Ptr0twistingBD}
  P_{\nu^\mathrm{L}_\beta \to \nu^\mathrm{R}_\alpha}(t) = &
  \frac{\sin^2 (2\theta)}{4}
  \bigg\{
    \left(
      \frac{\mu_1 B}{\mathcal{M}_1}\sin\mathcal{M}_1 t-
      \frac{\mu_2 B}{\mathcal{M}_2}\sin\mathcal{M}_2 t
    \right)^2
	\notag
	\\
	& +
    4\frac{\mu_1 \mu_2 B^2}{\mathcal{M}_1\mathcal{M}_2}
	\sin\mathcal{M}_1 t \sin\mathcal{M}_2 t \sin^2[\Phi(k)]
  \bigg\}.
\end{align}  
Here $\Phi(k)=\delta m^2/[4(k+\omega/2)]$ is the oscillations phase which now 
depends on the frequency of the twisting magnetic field. The next order 
correction in $\mu$ can be found in Ref.~\cite{Dvo07}.

Now we study the opposite case -- the magnetic moments matrix in 
Eq.~\eqref{magmomme} with great non-diagonal elements, $\mu \gg \mu_a$. This 
situation should be analyzed non-perturbatively. With help of the results of 
our work~\cite{Dvo07} we get the following transition probability:
\begin{align}\label{PtrtwistingBM}
  P_{\nu^\mathrm{L}_\beta \to \nu^\mathrm{R}_\alpha}(t)= &
  (\mu B)^2
  \left[
    \cos^2\theta \frac{\sin \varOmega_{+{}} t}{\varOmega_{+{}}}-
    \sin^2\theta \frac{\sin \varOmega_{-{}} t}{\varOmega_{-{}}}  	
  \right]^2,
\end{align}
where $\varOmega_{\pm{}}=\sqrt{(\mu B)^2+[\Phi(k) \pm \omega/2]^2}$.

Eqs.~\eqref{Ptr0twistingBD} and~\eqref{PtrtwistingBM} allow one to calculate 
transition probabilities for spin-flavor oscillations of Dirac neutrinos when 
particles interact with the twisting -- or spiral undulator -- magnetic field 
and propagate along the undulator axis since $\mathbf{k}=(k,0,0)$ and 
$\mathbf{B}=B(0,\sin \omega x,\cos \omega x)$. It should be noted that 
Eqs.~\eqref{Ptr0twistingBD} and~\eqref{PtrtwistingBM} reproduce the case of the 
constant transversal magnetic field, i.e. Eqs.~\eqref{Ptr0constBD} 
and~\eqref{PtrconstBM}, if we set $\omega=0$ there.

\section{Conclusion}\label{CONCL}

We have summarized the results of our studies of the evolution of Dirac 
neutrinos in external electromagnetic fields. To describe the time evolution of 
the neutrinos system we have applied the recently developed approach (see 
Refs.~\cite{DvoMaa07,Dvo07}) which is based on the the exact solutions to the 
Dirac-Pauli equation in an external magnetic field with the given initial 
condition. 

First (Sec.~\ref{CONSTB}) we have studied the dynamics of two mixed Dirac 
neutrinos with arbitrary magnetic moments matrix in the constant transversal 
magnetic field. Using earlier obtained solutions to the Dirac-Pauli equation we 
derive the neutrino wave function exactly accounting for all neutrino magnetic 
moments [Eq.~\eqref{nualphaconstB}] and valid for arbitrary magnetic fields. 
Then we have appled this result for oscillations of neutrinos with various 
magnetic moments matrices and derived transition probabilities 
[Eqs.~\eqref{Ptr0constBD} and~\eqref{PtrconstBM}]. We have discussed another 
magnetic field configuration in Sec.~\ref{TWISTB}. The evolution of Dirac 
neutrinos in the twisting magnetic field has been studied there. On the basis 
of the Dirac-Pauli equation solution in this external magnetic field we have 
obtained the transition probabilities [Eqs.~\eqref{Ptr0twistingBD} 
and~\eqref{PtrtwistingBM}] for different magnetic moments matrices.

\subsection*{Acknowledgments}

The work has been supported by the Academy of Finland under the
contract No.~108875. The author is thankful to the Russian Science Support 
Foundation for a grant as well as to the organizers of the 14$^\mathrm{th}$ 
International Baksan School "Particles and Cosmology" for the invitation and 
financial support.

\end{document}